\documentclass[prl,twocolumn,tightenlines,footinbib]{revtex4-2}
\usepackage{bm,dcolumn,amsmath,graphicx,amsfonts,amssymb}

\usepackage{hyperref}
\usepackage{xcolor}
\usepackage{comment}
\definecolor{cite}{rgb}{0.,0.,0.9} 
\hypersetup{colorlinks,linkcolor={cite},citecolor={cite},urlcolor={cite}}

\usepackage[capitalise]{cleveref}
\usepackage{siunitx}

\renewcommand{\vec}[1]{\ensuremath{{\boldsymbol{#1}}}}
\newcommand{\abs}[1]{\ensuremath{\left |#1\right |}}

\newcommand{\be}{\begin{equation}}	
\newcommand{\ee}{\end{equation}}	

\usepackage[normalem]{ulem}
\definecolor{newc}{rgb}{0.,0.6,0.4}

\renewcommand{\section}[1]{\vspace{0.85pt}\paragraph*{\textbf{\textit{\small{#1---}}}}}
\renewcommand{\subsection}[1]{\paragraph*{{\textit{\small{#1---}}}}}

\newcommand{\A}{\ensuremath{\mathcal{A}}} 

\begin{document}

\title{Empirical determination of the Bohr-Weisskopf effect in cesium and improved tests of precision atomic theory in searches for new physics}

\author{G.\ Sanamyan}\email[]{g.sanamyan@uq.edu.au}
\author{B.\ M.\ Roberts}\email[]{b.roberts@uq.edu.au}
\author{J.\ S.\ M.\ Ginges}\email[]{j.ginges@uq.edu.au}
\affiliation{School of Mathematics and Physics, The University of Queensland, Brisbane QLD 4072, Australia}
\date{\today}

\begin{abstract}

The finite distribution of the nuclear magnetic moment across the nucleus gives a contribution to the hyperfine structure known as the Bohr-Weisskopf (BW) effect. We have obtained an empirical value of $-0.24(18)\%$ for this effect in the ground and excited $s$ states of atomic $^{133}$Cs. This value is found from historical muonic-atom measurements in combination with our muonic-atom and atomic many-body calculations. The effect differs by $0.5\%$ in the hyperfine structure from the value found using the uniform magnetization distribution, 
which has been commonly employed in the precision heavy-atom community over the last several decades. 
We also deduce accurate values for the BW effect in other isotopes and states of cesium. These results enable cesium atomic wave functions to be tested in the nuclear region at an unprecedented $0.2\%$ level, and are needed for the development of precision atomic many-body methods. This is important for increasing the discovery potential of precision atomic searches for new physics, in particular for atomic parity violation in cesium. 

\end{abstract}

\maketitle

Atomic parity violation (APV) studies provide a low-energy precision means of 
interrogating the standard model of particle physics~\cite{GingesReview2004,RobertsReview2015,SafronovaReview2018,Zyla2020}. 
The dominant effect arises from Z-boson exchange between atomic electrons and the nucleus -- quantified by the nuclear weak charge $Q_W$ -- and the combination of experiment and atomic theory allows the value of $Q_W$ to be determined. The nuclear weak charge is uniquely sensitive to possible new physics such as Z’ and dark Z bosons~\cite{SafronovaReview2018}. The most precise APV measurement to date has been performed with $^{133}$Cs on the $6s$-$7s$ electric dipole transition with an uncertainty of $0.35\%$~\cite{Wood1997}, and an experimental program underway at Purdue~\cite{Toh2019} aims to improve upon this result. 
Atomic theory uncertainty has reached  sub-$0.5\%$~\cite{GingesCs2002,FlambaumQED2005,Porsev2009,*Porsev2010,OurCsPNC2012}, and there are efforts to reduce this towards $0.1$ -- $0.2\%$~\cite{Roberts2021,Trantan2022,Sahoo2021,RobertsComment2022}.

Critical to the testing and development of atomic many-body theory is the availability of precision experimental data to benchmark against. For APV calculations, which rely on high-accuracy modeling of the atomic wave functions across the extent of the atom, the hyperfine structure, transition energies, and usual electric dipole amplitudes provide these checks~\cite{GingesReview2004,GingesCs2002,Porsev2009,*Porsev2010}. Comparison between theory and experiment for the hyperfine structure tests the modeling of the wave functions in the nuclear region, where the weak interaction acts. While there is particularly high-quality experimental data available for cesium (the ground state hyperfine splitting in $^{133}$Cs defines the SI unit for time, the second~\cite{Heavner2014}), the atomic theory result relies on a model for the finite distribution of the nuclear magnetic moment across the nucleus, which cannot currently be accurately determined from nuclear structure theory. There is a high sensitivity to the choice of this magnetization model, and the hyperfine structure can vary by as much as $0.5\%$ for $^{133}$Cs~\cite{Ginges2017}. 
This uncertainty sets the lower limit on the uncertainty that can be claimed in atomic calculations, and the advancement of state-of-the-art atomic theory hinges on reducing this severalfold.

The finite-nucleus magnetization effect -- the Bohr-Weisskopf (BW) effect~\cite{Bohr1950,Bohr1951} --  also plays an important role in nuclear structure studies and tests of quantum electrodynamics (QED). It may be used for accurate determination of nuclear magnetic moments of short-lived isotopes~\cite{Persson2013,Prosnyak2020,Barzakh2020,RobertsFr2020,Konovalova2020}, and to probe the neutron distribution in atomic nuclei~\cite{Grossman1999,Zhang2015}. 
Uncertainties in its modeling from nuclear structure calculations obscure the QED contributions to the hyperfine structure in highly-charged ions, and its careful removal is required in tests of QED~\cite{Shabaev2001,Volotka2013,Skripnikov2018}. A proposal to remove the BW effect in many-electron atoms using measurements for high-lying states has been put forward~\cite{Ginges2018}, and new experimental data for $^{133}$Cs have been obtained~\cite{Quirk2022} that will allow this to be explored. 

It has been common practice over the last decades to use a uniform or Fermi distribution to model the nuclear magnetization in hyperfine structure calculations for heavy atoms (see, e.g., Ref.~\cite{Dzuba1984}). However, the more well-motivated single-particle (SP) model gives hyperfine structure values that may differ significantly: for $^{133}$Cs, the uniform distribution gives a contribution of $-0.7\%$ to the hyperfine structure, while the SP model yields only $-0.2\%$~\cite{Ginges2017}; the difference may be more than 1\% for other atoms of interest for precision tests of fundamental physics~\cite{Ginges2017}.   
The validity of the SP model for cesium isotopes -- and for several other atoms of interest for precision atomic searches for new physics, including francium and thallium -- is supported by measurements of differential hyperfine anomalies~\cite{Persson2013}, considered recently~\cite{Grossman1999,Zhang2015,Konovalova2018,Roberts2021,RobertsFr2020,Prosnyak2020,Prosnyak2021}. 
While this probes the {\it difference} in the finite-nucleus magnetization effects between different isotopes, we need the BW effect for a {\it single} isotope.

The answer lies in historical experimental data~\cite{Lee1969} on the hyperfine structure in muonic $^{133}$Cs. 
Due to the muon's proximity to the atomic nucleus, muonic atoms are highly sensitive to details of nuclear structure, including the nuclear charge radius and the BW effect~\cite{Wheeler1953,Engfer1974,Borie1982,Buttgenbach1984}. 
Experimental studies of BW effects in muonic atoms were superseded two decades ago by precision measurements in H-like ions, with a focus on systems of interest for QED tests~\cite{Klaft1994,Crespo1996,Crespo1998,Seelig1998,Beiersdorfer2001}. There is a new muonic-atom experimental program at Paul Scherrer Institut with a main goal to accurately determine the nuclear charge radius in $^{226}$Ra 
to support APV calculations for singly-ionized Ra~\cite{Knecht2020}; see also, e.g., the recent work with muonic rhenium~\cite{Antognini2020}.

In this work, we show how a BW effect in a muonic atom can be translated to that in a many-electron atom. 
First, we extract the BW effect from muonic $^{133}$Cs measurements and convert this to a BW value for the H-like ion, using a procedure introduced in Ref.~\cite{Elizarov2006}. This enables us to check our result by using the same method for $^{208}$Bi and $^{203,205}$Tl for which there is both muonic-atom and H-like data. We then use a recently-proposed method~\cite{Roberts2021scr} (see also Ref.~\cite{Skripnikov2020}), to convert a BW value in a H-like ion to that in a many-electron atom, to obtain accurate BW effects for any state of neutral $^{133}$Cs. Values for other isotopes are found from experimental data on  differential anomalies.

The muon in a muonic atom is located at a radius about 207 times closer to the nucleus than the electron for the corresponding state, owing to its larger mass. Therefore, it is largely unscreened by atomic electrons, and may be treated as a heavy H-like ion. Its wave functions may therefore be found from the Dirac equation,  
\be
    \big( c \vec{\alpha} \cdot \vec{p} + \left( \beta - 1 \right) m_\mu c^2 + V_\text{nuc}(r) \big) \phi_{n \kappa m} = \varepsilon \phi_{n \kappa m} \, , 
    \label{eqn:Dirac}
\ee
where $c$ is the speed of light, $\vec{\alpha}$ and $\beta$ are Dirac matrices, $m_\mu\approx 207\, m_e$ is the muon mass, $V_\text{nuc}(r)$ is the nuclear potential, and $\varepsilon$ is the binding energy. The indices $n$, $\kappa$, $m$ are the principal, relativistic angular, and magnetic quantum numbers, respectively. Unless otherwise specified, atomic units $\hbar = m_e = \abs{e} = c \alpha = 1$ are used throughout. The nuclear potential $V_{\rm nuc}$ is taken to correspond to a Fermi charge distribution, with a thickness of $\SI{2.3}{\femto\metre}$ and the root-mean-square radii $r_{\rm rms}$ taken from Ref.~\cite{Angeli2013}. 
The wave functions are expressed as 
\be
    \phi_{n \kappa m}({\bf r}) = \frac{1}{r} \begin{pmatrix} f_{n \kappa} (r) \Omega_{ \kappa m } \left( \vec{n} \right) \\
                                                  i g_{n \kappa} (r) \Omega_{-\kappa m } \left( \vec{n} \right) \end{pmatrix},
    \label{eqn:bispinor}
\ee 
where $f$ and $g$ are large and small radial components normalized as $\int_0^\infty \left( f^2 + g^2 \right)dr = 1$, $\Omega$ are spherical spinors, and $\vec{n} = \vec{r}/r$.

The magnetic hyperfine structure in a muonic atom arises from the magnetic interaction of the muon with the nuclear magnetic moment. 
The relativistic expression for this interaction may be written as  
\be
    h_{\text{hfs}} = \alpha  \vec{\mu} \cdot \left( \vec{r} \times \vec{\alpha} \right) F(r)/r^3, 
    \label{eqn:hfs}
\ee
where $\vec{\mu} = \mu \vec{I} / I$ is the nuclear magnetic moment, $\vec{I}$ is the nuclear spin, and $F(r)$ describes the distribution of the nuclear magnetic moment across the nucleus (with $F(r)=1$ for the point-nucleus case). 
We take nuclear magnetic moment values for $^{133}$Cs and $^{203,205}$Tl from Ref.~\cite{Stone2005}, and that for $^{209}$Bi from Ref.~\cite{Skripnikov2018}.   
The magnetic hyperfine structure is often quantified by the hyperfine constant $\A$, which is found by averaging the interaction \cref{eqn:hfs} over the atomic state. For the case of point-nucleus magnetization, and in the $1s$ state of the muonic atom, the hyperfine constant is given by 
\be
    \A_0 =  \frac{4}{3} \frac{\alpha}{I} \frac{1}{m_p}  \frac{\mu}{\mu_N} \int_0^{\infty} dr f(r) g(r) / r^2, 
    \label{eqn:pointlikedE}
\ee
where $\mu_N$ is the nuclear magneton and $m_p$ the proton mass. Finite-nuclear-charge effects are included through $f$ and $g$, which are found in the nuclear Coulomb potential of a finite charge distribution (see \cref{eqn:Dirac}).

The total hyperfine constant may be expressed as  
\be
    \A = \A_0 + \A_{\text{BW}} + \A_{\text{QED}} \, ,
    \label{eqn:totaldE}
\ee
where $\A_{\text{BW}}$ is the Bohr-Weisskopf effect, which gives the finite-nucleus magnetization contribution, and $\A_{\text{QED}}$ is the QED correction. 
In muonic atoms, the BW effect may enter at the level of $100\%$ of the hyperfine constant, and the QED correction is much smaller. This is strikingly different to the case for usual atoms and ions, where the BW effect is typically several $0.1\%$ to several $1\%$, and the sizes of the BW and QED corrections are comparable~\cite{LeBellac1963,Buttgenbach1984,Shabaev1994,Elizarov2006}.

\begin{table*}[tb]
    \caption{Contributions to $1s$ hyperfine constants of muonic atoms (point-nucleus results $\A_0$, QED contributions $\A_{\rm QED}$, measured values $\A_{\rm exp}$) and extracted BW effects $\A^{\rm exp}_{\rm BW}$ compared to predictions of nuclear magnetization models -- uniform distribution ``ball", single-particle ``SP", SP with Woods-Saxon potential and spin-orbit interaction ``SP-WS", configuration mixing ``CM", and microscopic theory ``FA$_{\rm I}$" and ``FA$_{\rm II}$" with two different sets of nuclear parameters. Units: $\SI{}{\kilo\electronvolt}$.}
    \label{tbl:extracted_A}
    \begin{ruledtabular}
    \begin{tabular}{ccccccccccc}
      & $\A_0$ & $\A_\text{QED}$ & $\A_{\rm exp}~$\cite{Buttgenbach1984} & $\A^{\rm exp}_\text{BW}$ & $\A_\text{BW}^{\text{ball}}$ & $\A_\text{BW}^\text{SP}$ & $\A_\text{BW}^{\rm SP-WS}$ & $\A_\text{BW}^\text{CM}$~\cite{Johnson1970} & $\A_\text{BW}^{\text{FA}_\text{I}}$~\cite{Fujita1975}& $\A_\text{BW}^{\text{FA}_\text{II}}$~\cite{Fujita1975}\\
    \hline
    $^{133}$Cs & $0.762$ & $0.006(3)$   & $0.634(103)$\tablenotemark[1] & $-0.134(103)$ & $-0.315$ & $-0.118$ & $-0.118$ & $-0.074$ & $-0.100$ & $-0.124$ \\
    $^{203}$Tl & $4.712$ & $0.023(12)$   & $2.340(80)$ & $-2.395(80)$ & $-2.481$ & $-2.481$ & $-2.069$ & $-1.688$ & $-1.974$ & $-2.125$ \\
    $^{205}$Tl & $4.744$ & $0.023(12)$   & $2.309(35)$ & $-2.458(37)$ & $-2.499$ & $-2.499$ & $-2.082$ & $-1.664$ & $-2.016$ & $-2.177$ \\
    $^{209}$Bi & $1.339$ &  $0.006(3)$   & $0.959(52)$ & $-0.386(52)$ & $-0.714$ & $-0.415$ & $-0.464$ & $-0.436$ & $-0.356$ & $-0.420$ \\
    \end{tabular}
    \end{ruledtabular}
    \tablenotetext[1]{Reference~\cite{Lee1969}. In the review~\cite{Buttgenbach1984} the uncertainty is erroneously presented a factor of 10 too small.}
\end{table*}

The QED radiative corrections to the hyperfine constant for muonic atoms are dominated by the vacuum polarization~\cite{Elizarov2006},
which consists of two parts, electric loop $\A_{\text{VP}}^{\text{EL}}$ and magnetic loop $\A_{\text{VP}}^{\text{ML}}$, 
\be
    \A_{\text{QED}} \approx \A_{\text{VP}}^{\text{EL}} + \A_{\text{VP}}^{\text{ML}}\, .
    \label{eqn:dEqed_total}
\ee 
We evaluate these contributions in the Uehling approximation~\cite{Uehling1935}, and account for finite-nucleus charge and magnetization distributions. 
The electric loop contribution is found by adding the Uehling potential to the nuclear potential in the Dirac equation, \cref{eqn:Dirac}, and finding the correction to the hyperfine constant with the perturbed wave functions. Finite-nucleus expressions for the Uehling potential may be found, e.g., in Refs.~\cite{Artemyev2001,Ginges2016}. 
The magnetic loop correction is evaluated through the introduction of an operator~\cite{Volotka2008,Schneider1994,Artemyev2001}, and we use the finite-nucleus expression in Ref.~\cite{Volotka2008}. The electric and magnetic vacuum polarization corrections are of comparable size, and account of finite-nucleus magnetization is important, reducing the point-nucleus values by factors of $2$ -- $3$. Overall, however, the QED contributions are small. Our results are presented in \cref{tbl:extracted_A}, with a conservative uncertainty estimate of 50\% the size of the contribution.

With accurate theoretical evaluations for $\A_0$ and $A_{\rm QED}$, empirical values for the BW effect $A_{\rm BW}^{\rm exp}$ may be extracted from the measured hyperfine constants,
\begin{equation}
\A_{\rm BW}^{\rm exp}= \A_{\rm exp}-\A_0-\A_{\rm QED}  \ .
\label{eqn:empiricalBW}
\end{equation}
Our empirical BW results are presented in \cref{tbl:extracted_A}. 
Our values for Tl and Bi are similar to those determined previously~\cite{Buttgenbach1984,Elizarov2006}, while for Cs the previous result~\cite{Buttgenbach1984} contained a factor of $10$ error in the uncertainty. 
These results allow the validity of different nuclear magnetization models to be tested, and in \cref{tbl:extracted_A} we give the BW values predicted by three simple models, as well as from more sophisticated calculations~\cite{Johnson1970,Fujita1975}. 
The three models considered are: (i) uniform magnetization distribution (ball model), $F(r) = (r/R_m)^3$ within $r<R_m$, that has been commonly used in heavy-atom calculations of the hyperfine structure; (ii) single-particle (SP) model, with the nucleon probability density taken to be constant across the nucleus, utilized in calculations of heavy hydrogen-like ions and heavy atoms, see, e.g., Refs.~\cite{Shabaev1994,Volotka2008}; and (iii) SP model with spin-orbit interaction taken into account, and the nucleon wave function found in the Woods-Saxon potential~\cite{Rost1968, Shabaev1997}, ``SP-WS". 
Calculations of the BW effect are obtained from
\be
    \A_{\text{BW}} = \frac{4}{3} \frac{\alpha}{I} \frac{1}{m_p} \frac{\mu}{\mu_N} \int_0^\infty dr \left( F(r) - 1 \right) f(r) g(r) / r^2. 
    \label{eqn:get_BW}
\ee 
It is often more convenient to consider the relative correction $\epsilon$, 
\be
\epsilon = \A_\text{BW}/\A_0  \, .
\ee
It is seen from \cref{tbl:extracted_A} that for $^{133}$Cs and $^{209}$Bi the ball model is a grossly inadequate representation, while the nuclear SP model -- and configuration mixing~\cite{Johnson1970} and microscopic~\cite{Fujita1975} calculations -- gives results in reasonable agreement with experiment. This is in line with our recent observations that for differential hyperfine anomalies~\cite{Roberts2021}, the SP model on the whole outperforms the ball model across the board. 

\begin{figure}[bb]
    \centering
    \includegraphics[width=.45\textwidth]{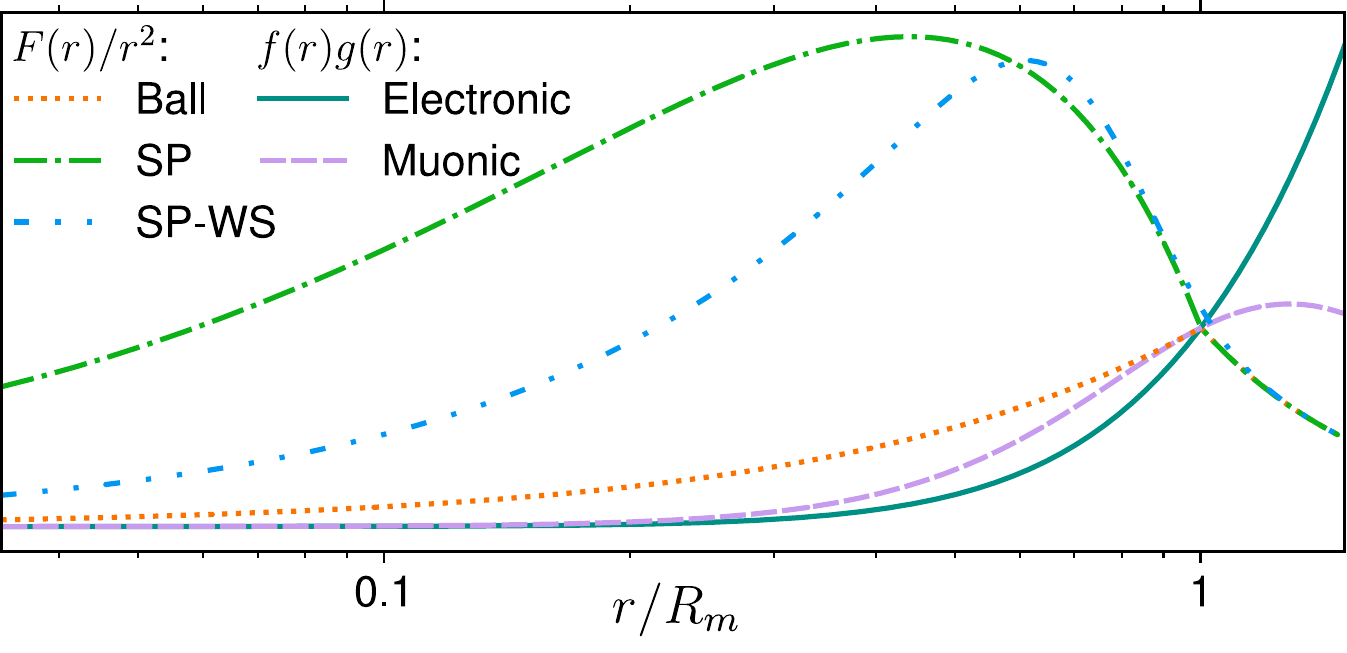}
    \caption{Product of $1s$ radial wave functions  $fg$ for H-like and muonic  $^{133}$Cs as a function of radial distance $r/R_m$, where $R_m$ is  nuclear magnetic radius, alongside $F(r)/r^2$ that describes the nuclear magnetization distribution in ball, SP, SP-WS models.  
    Values $fg$ are scaled to be unity at $r=R_m$.}
    \label{fig:plot_sols}
\end{figure}

We now proceed to an empirical result for the BW effect in atomic Cs from the experimentally-deduced BW effect in muonic Cs, $\A_{\rm BW}^{\rm exp}$. Translating the result from muonic to electronic atoms is not completely straightforward. In usual atoms, the wave functions for $s$ states are proportional in the nuclear region, independent of the principal quantum number and ionization degree (see, e.g., Refs.~\cite{Shabaev2001,FlambaumQED2005,Ginges2018}). This proportionality for states of a hydrogen-like ion and neutral atom allowed us~\cite{Roberts2021scr} to determine accurate and model-independent electronic screening factors, that may be applied, e.g., to empirically-deduced BW effects from H-like ions for use in neutral atom calculations. We will use this in the second step of a two-step process, first obtaining an empirical result for H-like cesium. 

In \cref{fig:plot_sols}, products of wave functions of the ground states for muonic and H-like $^{133}$Cs are presented. 
It is seen that these have a different form in the nuclear region, giving these systems a different sensitivity to the nuclear magnetization distribution across $r$. To more clearly illustrate this, the function $F(r)/r^2$, in the three simple models considered above, is shown alongside.

\begin{table}[tt]
    \caption{Relative BW corrections $-\epsilon$ ($\%$) to lowest $s$ states of muonic atoms, H-like ions, and neutral Cs.  
    Subheadings ``$\mu$ exp." and``H-like exp." indicate that results are extracted from muonic and H-like experiments, respectively.} 
    \label{tbl:results_fitted}  
    \begin{ruledtabular}
    \begin{tabular}{ccccccc}
      & $\mu$-atoms && \multicolumn{2}{c}{H-like ions} && \multicolumn{1}{c}{atoms} \\
      \cline{2-2}\cline{4-5}\cline{7-7}
      & $\mu$ exp. && $\mu$ exp. & H-like exp. && $\mu$ exp.  \\ 
    \hline
    $^{133}$Cs & $18(14)$ && $0.23(17)$  & $-$                      && $0.24(18)$ \\   
    $^{203}$Tl & $50.8(1.6)$ && $1.93(15)$ & $2.21(8)$\tablenotemark[1] && $-$  \\
    $^{205}$Tl & $51.8(8)$   && $1.98(15)$ & $2.25(8)$\tablenotemark[1] && $-$  \\
    $^{209}$Bi & $28.8(3.9)$ && $0.98(14)$ & $1.03(5)$\tablenotemark[2] && $-$  \\
    \end{tabular}
    \end{ruledtabular}
    \tablenotetext[1]{Reference \cite{Beiersdorfer2001};
    $^{\rm b}$\,Reference \cite{Ullmann2017}.}
\end{table}

To translate the empirical BW effect in a muonic atom to a BW effect for the H-like system, and account for the nuclear model dependence, we follow the procedure introduced in Ref.~\cite{Elizarov2006}. We use the simple SP magnetization models from that work, and find effective nuclear magnetic radii $R_m$ such that the experimental BW effects in muonic atoms are reproduced.  
These SP models have different parametrizations for the unpaired nucleon wave function across the nucleus: 
zeroth, first, and second powers  
of $r$ and $(R_m-r)$. These models are then applied to the H-like systems, and BW effects determined. The uncertainty due to the model dependence is captured through the spread in results, and the central value is given by the midpoint. 
For H-like $^{133}$Cs we obtain $\epsilon = -0.227(174)(10)\%$, where the first uncertainty is experimental and the second is due to the model dependence.

Our final results are presented in \cref{tbl:results_fitted}. The uncertainties for the H-like results determined from muonic atom experiments comprise of those from 
experiment (the relative uncertainty is taken to be equal to that for the muonic atom value $\rm \A^{\rm exp}_{BW}$) and those from the model dependence of the magnetization distribution, added in quadrature. 
For Cs and Bi, the former dominates, while for Tl the latter is larger by $2$ -- $3$ times. The results for Tl$^{80+}$ and Bi$^{82+}$ may be checked against accurate data from direct H-like experiments, shown in the table. It is seen that the data for Bi$^{82+}$ are in agreement, and for Tl$^{80+}$ the results lie only marginally outside the $1\sigma$ error bars. This gives confidence in the result for Cs$^{54+}$, and that any missed effects, such as from nuclear polarization (considered, e.g., for the hyperfine structure in muonic deuterium in Ref.~\cite{kalinowski2018} and for energies of heavy muonic atoms in Ref.~\cite{Valuev2022}), may be omitted at the current level of uncertainty.

In the final column of \cref{tbl:results_fitted} we present the result for the ground state of $^{133}$Cs. This is found using an electronic ``screening" factor~\cite{Roberts2021scr}, 
which relates the BW effect in a many-electron atom or ion to the BW effect in the corresponding H-like ion, $x_{\rm scr}(6s)= 1.047$. 
Our empirically-derived BW effect for the $6s$ state of atomic $^{133}$Cs, $-0.24(18)\%$,
agrees with nuclear single-particle predictions, with the SP value $-0.21\%$ \cite{Ginges2017,Roberts2021} and the SP-WS value $-0.19(14)\%$~\cite{Ginges2017}, and with configuration mixing calculations, $-0.22\%$~\footnote{Obtained from the configuration mixing calculation for $^{137}$Cs in combination with calculated differential anomalies, Ref.~\cite{Stroke1961}.}. 
However, it deviates significantly from predictions ($-0.7\%$) of the commonly-used uniform distribution, amounting to a sizeable $0.5\%$ difference in the hyperfine structure constant. 
Taken together with the high-precision many-body result of Ref.~\cite{Ginges2017} for the ground-state hyperfine structure constant for $^{133}$Cs, our empirical result corresponds to an absolute BW correction of $-5.5(4.2)\,{\rm MHz}$.   
This yields a total value of $2293.1\,{\rm MHz}$, which agrees with experiment ($2298.157...$\,MHz~\cite{RevModPhys.49.31}) at the level of the uncertainty from the BW effect.

\begin{table}
    \caption{Relative BW corrections $\epsilon^{(2)}_{s}$ for $s$ states of atomic cesium isotopes, from measured differential anomalies $^1 \Delta^2_\text{exp}$, our BW value for $^{133}$Cs, $\epsilon^{(1)}= -0.24(18)\%$, and calculated differential BR effects $^{1} \delta^{2}$. $A^{(1)}$, $A^{(2)}$  and $I^\pi_1$, $I^\pi_2$ are atomic mass numbers and nuclear spins of isotopes 1 and 2.}
    \label{tbl:HFS_CS_iso}
    \begin{ruledtabular}
    \begin{tabular}{ccccccc}
    $A^{(1)}$ & $I^\pi_1$ & $A^{(2)}$  & $I^\pi_2$ & $^{1} \Delta^{2}_\text{exp}$ ($\%$) & $^{1} \delta^{2}$ ($\%$) & $-\epsilon^{(2)}_{s}$ ($\%$)\\
    \hline

     $133$ & $7/2^+$ & $131$ & $5/2^+$ & $0.45(5)$\tablenotemark[1]   & $-0.001$ & $0.69(19)$ \\
     $133$ & $7/2^+$ & $134$ &   $4^+$ & $0.169(30)$\tablenotemark[2] &  $0.000$ & $0.41(18)$ \\
     $133$ & $7/2^+$ & $135$ & $7/2^+$ & $0.037(9)$\tablenotemark[3]  &  $0.001$  & $0.27(18)$ \\
     
    \end{tabular}
    \end{ruledtabular}
    \tablenotetext[1]{Reference \cite{Worley1965}; 
    $^{\rm b}$\,Reference \cite{Persson2013};
    $^{\rm c}$\,Reference \cite{Stroke1957}.}
\end{table}

The BW effects in $s$ and $p$ states exhibit weak dependence on principal quantum number~\cite{Perez2007,Ginges2018,Grunefeld2019}, and the obtained result for $6s$ is valid also for excited $ns$ states of $^{133}$Cs. Furthermore, since the BW effect in atomic states may be determined from the $1s$ state in the corresponding H-like ion~\cite{Roberts2021scr} (see also~\cite{Skripnikov2020}), the effect in other states may also be found. 
We find the BW effects $-0.015(12)\%$ for   $np_{1/2}$ states and $-0.065(50)\%$ for  $np_{3/2}$ states.

We use the empirical BW result for $^{133}$Cs to obtain values of the BW effects for other isotopes using measured data for differential hyperfine anomalies. The differential anomaly ($^1 \Delta^2$) between isotopes $1$ and $2$ is defined as~\cite{Persson2013}
\be
    ^1 \Delta^2 = (\A_1/\A_2) (g_2/g_1) -1 \approx {^1 \delta^2} + \epsilon^{(1)} - \epsilon^{(2)},
    \label{eqn:differHFanomal}
\ee
where $^1 \delta^2=\delta^{(1)} - \delta^{(2)}$ is the differential Breit-Rosenthal (BR) effect (modeled accurately using a Fermi charge distribution for $V_\text{nuc}(r)$ in \cref{eqn:Dirac}), and $\epsilon^{(1)}$ and $\epsilon^{(2)}$ are BW effects for isotopes $1$ and $2$. The results for $s$ states for isotopes $A=131,\, 134, 135$ are presented in \cref{tbl:HFS_CS_iso}.

In summary, we have determined accurate empirical Bohr-Weisskopf effects for $s$ and $p$ states of Cs isotopes. For $s$ states, the associated uncertainty is within 0.2\% the size of the hyperfine constants. This resolves the 0.5\% tension between predictions of different nuclear magnetization models in favor of the nuclear SP model. 
Our results may be used to correct previous calculations of the hyperfine structure, and we recommend their adoption in future theoretical evaluations. They have implications for the error analyses of the most precise determinations~\cite{GingesCs2002,Porsev2009,*Porsev2010} of the $^{133}$Cs atomic parity violation amplitude, with claimed uncertainties $0.27\%-0.5\%$. Indeed, the hyperfine constants for the $6s$ and $7s$ states should be corrected by $0.4\%-0.5$\% from the Fermi or uniform magnetization distributions used in those works; the $p_{1/2}$ states are unaffected due to the order-of-magnitude smaller relative BW corrections. This leads to a shift of $0.2\%-0.3\%$ in the values for $\sqrt {\A_{ns}\A_{np_{1/2}}}$, which is considered to give an indication of the accuracy of the $ns-np_{1/2}$ weak matrix elements. The effect is an increase by 0.4\% in the deviation from experiment for the hyperfine constant for the $6s$ state in Ref.~\cite{Porsev2009,*Porsev2010}, and an increase by 0.2\% for the associated square root formula, while the results of Ref.~\cite{GingesCs2002} remain essentially unchanged~\footnote{Due to the competing contribution of QED effects that was unaccounted for in that work.}.

The results of the current work allow testing of the cesium atomic wave functions in the nuclear region at an unprecedented level of 0.2\%, paving the way for next-generation precision atomic many-body calculations. The uncertainties of the results are overwhelmingly experimental, and we emphasize the importance of new, more precise experiments with muonic atoms, or with simpler systems such as H-like ions with cleaner theoretical interpretation, for further reducing uncertainties in the BW effect to sub-0.1\% for cesium and other systems of interest for precision atomic searches for new physics.

\section{Acknowledgments}
We are grateful to N. Oreshkina for useful discussions. This work was supported by the Australian Government through Australian Research Council (ARC) DECRA Fellowship No.\ DE210101026 and ARC Future Fellowship No.\ FT170100452.

\bibliography{library}
\end{document}